\documentstyle[11pt,epsfig]{article}

\begin{document}

\begin{titlepage}
\begin{flushright}
CERN-PH-TH/2005-115
\end{flushright}

\vspace{0.8cm}

\begin{center}

\huge{Gradient expansion(s) and dark energy}

\vspace{0.8cm}

\large{Massimo Giovannini \footnote{e-mail address: massimo.giovannini@cern.ch}}

\normalsize
\vspace{0.3cm}
{{\sl Centro ``Enrico Fermi", Compendio del Viminale, Via 
Panisperna 89/A, 00184 Rome, Italy}}\\
\vspace{0.3cm}
{{\sl Department of Physics, Theory Division, CERN, 1211 Geneva 23, Switzerland}}
\vspace*{1cm}

\begin{abstract}
\noindent
Motivated by recent claims stating that the acceleration 
of the present Universe is due to fluctuations with 
wavelength larger than the Hubble radius, we present a general 
analysis of various  perturbative solutions of fully inhomogeneous Einstein 
equations supplemented by a perfect fluid. The equivalence of formally  different gradient expansions is demonstrated. 
If the barotropic index vanishes, the deceleration parameter is always positive semi-definite.
\end{abstract}
\end{center}

\end{titlepage}

\newpage
\renewcommand{\theequation}{1.\arabic{equation}}
\section{Introduction}
\setcounter{equation}{0}

Type Ia supernovae seem to suggest 
that the present Universe is experiencing a phase of accelerated 
expansion \cite{riess1,tonry1} (see also \cite{riess2,perlmutter} 
for earlier results reporting the first evidences of this phenomenon).
The experimental results are also consistent with the determination
of the cosmological parameters coming from CMB physics \cite{wmap}. 
These results seem to imply that up to 70 \% of the 
present energetic content of the Universe is ``dark".  
If the dark energy component is parametrized with a perfect fluid 
characterized by a barotropic index $w$, the experimental 
determinations reported in Refs. \cite{riess1,tonry1}  suggest 
that for a flat Universe with $\Omega_{\rm tot} = 1$, $\Omega_{\rm matter}
= 0.29 \pm^{0.05}_{0.03}$ and $w = -1.02\pm^{0.13}_{0.19}$.

The simplest model of dark energy is, in some sense, a 
time-independent cosmological constant whose associated $w$ is 
exactly $-1$.  Fluid models of dark energy may lead to barotropic
indices in the range $-1 \leq w\leq -1/3$.  It is equally plausible 
that the dark-energy component is described by a scalar degree 
of freedom (i.e. the quintessence field) whose potential becomes dominant around the present time \cite{q1} (see also \cite{q1a}).
 Such a degree of freedom may also be a pseudo-scalar \cite{q2}. Quintessence models of dark energy 
imply generically that $w > -1$. A rather economical class of models 
 is the one where the quintessence field (dominant today) 
and the inflaton field (dominant in the far past history of the Universe)
are identified in a single (scalar) degree of freedom. These are the so-called 
quintessential inflationary models \cite{q3} whose salient features 
may lead necessarily to a stochastic background of gravitational radiation
for typical frequencies larger than $0.1 \,\,{\rm KHz}$  \cite{mg1,mg2,mg3}.
Useful surveys on the status of fluid and quintessential models 
of dark energy can be found, respectively, in Refs.  \cite{padma} and \cite{ratra}.

Experimental data cannot rule out  values $ w< -1$: this 
occurrence stimulated the investigation of fluid models of dark energy where 
the cosmological energy density is future increasing, rather than 
future decreasing, as in the case $w >-1$. Provided 
$w < -1$ all the time in the future, a ``big rip" can be expected \cite{rip}, i.e.
an infinite expansion of the Universe in a finite amount of time 
\cite{rip2,visc1}. Along this perspective, recently Barrow  has 
discussed the interesting case  of ``sudden" singularities that may 
 arise in expanding FRW Universes even if the dominant energy condition is not violated \cite{barrow1,barrow2}. The possibility of anisotropic 
 sudden singularities has been also discussed recently by Barrow and Tsagas
 \cite{barrow3}. We will briefly comment about the possible relevance 
 of the considerations developed here in the context of the physics 
 of sudden singularity.

The aim of the present paper is to understand wether an inhomogeneous 
Universe may sustain a phase of accelerated expansion even if 
the barotropic index of the sources is positive semi-definite, i.e. $w \geq 0$.
There are various reasons to discuss this problem. It is, in fact, 
plausible that inhomogeneities (with typical wavelengths 
larger than the Hubble radius) may be generated in the 
early Universe provided the duration of the inflationary phase 
exceeds, say, $65$ e-folds.  After equality and before decoupling,  
(when inflationary perturbations are imprinted, via the Sachs-Wolfe 
effect, on the temperature anisotropies) the metric fluctuations 
are perturbative.  Still, one could wonder if higher-order effects 
in the amplitude of the fluctuations may change this perspective and make 
the present amplitude of inflationary generated fluctuations so large 
to affect, for instance, the deceleration parameter. 

There may be a way out in this technical impasse. The idea 
is to forget about homogeneous backgrounds (supplemented 
by tiny inhomogeneous fluctuations) and to consider, instead, some 
type of gradient expansion. This theoretical tool has a 
long history that can be traced back to the seminal 
contributions of Lisfhitz, Belinskii, Khalatnikov and collaborators 
(see \cite{LK1,BLK} and references therein). 
This approach allows to compute the 
phenomenologically relevant quantities to a given order in the spatial
gradients. There have been recently claims \cite{K}  
 suggesting that indeed a matter-dominated 
Universe (without any vorticity) may be accelerating 
solely thanks to the presence of super-Hubble inhomogeneities \footnote{To stress the fact that the Universe is accelerated thanks to super-horizon 
fluctuations, the authors of \cite{K} called their proposal Super Hubble 
Cold Dark Matter (SHCDM) to be contrasted with the usual $\Lambda$CDM.}. In   
\cite{GCA,Fla,selhir,alnes,RAS,siegel,mgio3}  various independent 
criticisms of this proposal have been presented.
In \cite{GCA} it was argued that fluctuations with typical 
wavelength larger than the Hubble radius can certainly 
affect the spatial curvature but can never accelerate the Universe.
Similar conclusion, through a different chain of arguments, has 
been derived in \cite{Fla}.
In \cite{selhir} demonstrated that super-horizon 
fluctuations do not produce an accelerated expansion. Their 
approach has been to take an exact solution of Einstein equations 
\footnote{Recently, a similar perspective has been adopted 
in the analysis of a Lema\^itre-Tolman-Bondi Universe by Alnes, Amarzguioui and Gron \cite{alnes} (see also \cite{ras} ).}
(describing an underdense region).  This strategy allowed to 
obtain a specific (computable) model of super-horizon 
fluctuations falling in the class discussed in \cite{K}. The 
example showed that no acceleration takes place in agreement 
with general theorems implying that the Universe cannot accelerate
unless the strong energy condition is violated. In \cite{selhir} an explicit 
analysis of the variance of the deceleration parameter has been also performed in the case of single-field inflationary models. 
In \cite{RAS} the perspective has been to analyse a form of averaged 
Einstein equations by assuming the correctness of some of the 
conclusions of \cite{K}. The conclusion of this study has been that 
the model of \cite{K} is likely to be ruled out by present observations.

The analysis of Ref. \cite{siegel} dealt also with 
the effect of small-scale fluctuations (i.e. with wavelengths 
smaller than the Hubble radius). The conclusion of the analysis 
has been that sub-horizon perturbations are not a viable 
candidate for the explanation of the present acceleration of the Universe.

In \cite{mgio3}, an approach for the analysis of fully inhomogeneous 
Einstein equations  has been proposed and applied to the case 
of Ref. \cite{K}. The calculation of \cite{K} has been then repeated 
and it has been discovered that, within the accuracy of the solution, the 
late-time behaviour of the spatial gradients in a matter dominated 
Universe can never make the deceleration parameter negative.
Furthermore, by enforcing the validity of the perturbative expansion 
it has been shown that, at later times  the deceleration parameter vanishes, i.e. the expected result is the Universe is dominated by gradients.

The purpose of the present paper is to extend and generalize the analysis 
presented in \cite{mgio3} along different lines. The first generalization 
consist in discussing the gradient expansion without the restriction 
of a dust-dominated Universe but for a generic barotropic 
index. Furthermore, in \cite{mgio3} the analysis has been 
performed by only keeping only one scalar degree of freedom of the 
fully inhomogeneous metric. Here, as announced \cite{mgio3} 
all the degrees of freedom of the inhomogeneous metric will be analyzed simultaneously.  The related aim of the present analysis is to stress 
the mutual  connections of apparently diverse gradient expansions.

The plan of the present paper is the following. In Section 2 the fully
inhomogeneous Einstein equations will be introduced in terms 
of the extrinsic and intrinsic curvature tensors. Section 3 deals with
the interesting case of anti-newtonian solutions that could be an appropriate 
seed  solution for a gradient expansion aiming at the 
description of super-horizon fluctuations. In Section 4 the gradient expansion will be derived. Solutions accurate up to two gradients will be presented. In Section 5 the gradient expansion will be applied to a situation where 
the parametrization of the seed metric is apparently different. 
Section 6 contains some 
more phenomenological considerations dealing with the late-time behaviour 
of a Universe dominated by super-Hubble fluctuations. Finally,
Section 7 contains some concluding remarks.

\renewcommand{\theequation}{2.\arabic{equation}}
\section{Inhomogeneous Einstein equations}
\setcounter{equation}{0}
Consider a four-dimensional line element of the form 
\begin{equation}
ds^2 = dt^2 - \gamma_{ij}(t, \vec{x}) dx^{i} dx^{j};
\label{metric}
\end{equation}
the symmetric rank-two tensor $\gamma_{ij}$ in three dimensions 
 contains $6$ independent degrees of freedom. 
 The components of the extrinsic curvature are 
\begin{equation}
 K_{i j} = - \frac{1}{2} \frac{\partial}{\partial t} \gamma_{i j},\qquad K_{i}^{j} = - \frac{1}{2} \gamma^{i k} \frac{\partial }{\partial t} \gamma_{k j}.
\label{extr}
\end{equation}
In the following, 
the compact notation ${\rm Tr} K^2 = K_{i}^{j} \, K_{j}^{i}$ will be used.
The three-dimensional Ricci tensor becomes instead
\begin{equation}
r_{i j} = \partial_{m} \Gamma^{m}_{i j} - \partial_{j} \Gamma^{m}_{m i} + 
\Gamma_{i j}^{m} \Gamma_{m \ell}^{\ell} - \Gamma_{j m}^{\ell} \Gamma_{i \ell}^{m},
\label{intr} 
\end{equation}
where 
\begin{equation}
\Gamma_{i j}^{m}= 
\frac{1}{2} \gamma^{m \ell}( - \partial_{\ell} \gamma_{i j} +\partial_{j}\gamma_{\ell i} + \partial_{i} \gamma_{j \ell}).
\label{christ}
\end{equation}
Equations (\ref{extr}) and (\ref{intr}) allow to write the Einstein equations 
in a fully inhomogeneous form. More specifically, assuming 
that the energy-momentum tensor is a perfect 
relativistic fluid 
\begin{equation}
T_{\mu}^{\nu} = (p+ \rho) u_{\mu}\,u^{\nu} - p \delta_{\mu}^{\nu},
\end{equation}
the Hamiltonian and momentum constraints are, respectively,
\begin{eqnarray}
&& K^2 - {\rm Tr} K^2  + r  =  16\pi G [ ( p + \rho) u_{0} u^{0} - p],
\label{E100}\\
&& \nabla_{i} K - \nabla_{k} K^{k}_{i} = 8\pi G u_{i} u^{0} (p + \rho).
\label{E10i}
\end{eqnarray}
The $(ij)$ components of Einstein equations lead instead to
\begin{eqnarray}
&&( \dot{K}_{i}^{j} - K \, K_{i}^{j} - \dot{K} \delta_{i}^{j}) + \frac{1}{2} \delta_{i}^{j}( K^2 + {\rm Tr} K^2) - (r_{i}^{j} - \frac{1}{2} r \delta_{i}^{j})  
\nonumber\\
&&= - 8\pi G [ (p + \rho) u_{i} u^{j} + p \delta_{i}^{j} ],
\label{E1ij}
\end{eqnarray}
where the overdot denotes a 
 partial derivation with respect to $t$ while $\nabla_{i}$ denotes 
 the covariant derivative defined in terms of $\gamma_{ij}$ and of Eq. (\ref{christ}).
A trivial  remark is that, in Eqs. (\ref{E100}), (\ref{E10i}) 
and (\ref{E1ij}), the indices are raised and lowered using directly 
$\gamma_{ij}(t,\vec{x})$. 

By combining the previous set of equations the following relation  can be 
easily deduced
\begin{equation}
q\, {\rm Tr} K^2 = 8\pi G \biggl[(p + \rho) u_{0}u^{0} + \frac{p - \rho}{2} \biggr]
\label{00cont}
\end{equation}
 where 
\begin{equation}
q(\vec{x},t) =  - 1 + \frac{\dot{K}}{{\rm Tr} K^2},
\label{defq}
\end{equation}
is the inhomogeneous generalization of the deceleration parameter.
In fact, in the homogeneous and isotropic limit, $ \gamma_{ij} = a^2(t) 
\delta_{ij}$, $K_{i}^{j} = - H \delta_{i}^{j}$ and, as expected, $q(t) \to - \ddot{a} a/ \dot{a}^2$.
Recalling the definition of ${\rm Tr}K^2$ (given after Eq. (\ref{extr})) it is 
rather easy to show that
\begin{equation}
{\rm Tr} K^2 \geq \frac{K^2}{3} \geq 0,
\label{trin}
\end{equation}
where the sign of equality (in the first relation) is reached, again, in 
the isotropic limit.

Since $\gamma^{ij}$ is always positive semi-definite, it is
also clear that 
\begin{equation}
 u_{0}\,u^{0} = 1 + \gamma^{ij} u_{i} u_{j} \geq 1,
\label{u0in}
\end{equation}
that follows from the condition $g^{\mu\nu} u_{\mu} u_{\nu}=1$.
From Eq. (\ref{00cont}) it follows
that $q(t,\vec{x})$ is always positive semi-definite if $(\rho + 3p) \geq 0$.
As correctly pointed out in \cite{selhir} and \cite{RAS} 
that this statement assumes the absence of rotational fluctuations. 
This is rather plausible since super-horizon vector modes 
are very unlikely to be generated either during inflation 
or during a phase where the Universe expands.  Concerning the 
possible r\^ole of vector modes in conventional and unconventional 
inflationary models see, for instance, \cite{v1,v2,v3,v4}. 

Equations (\ref{E100}), (\ref{E1ij}) and (\ref{E10i}) must be supplemented 
by the explicit form of the covariant conservation equations:
\begin{eqnarray}
&& \frac{1}{\sqrt{\gamma}} \frac{\partial}{\partial t} [ \sqrt{\gamma} ( p + \rho)
u^{0} u^{i}] - \frac{1}{\sqrt{\gamma}} \partial_{k}\{ \sqrt{\gamma} [ (p+ \rho) 
u^{k} u^{i} + p\gamma^{ki}]\}  
 -2 K^{i}_{\ell} u^{0} u^{\ell} (p + \rho)
 \nonumber\\
&& - \Gamma_{k \ell}^{i} 
[ (p + \rho) u^{k} u^{\ell} + p \gamma^{k\ell}] =0,
\label{con1}\\
&& \frac{1}{\sqrt{\gamma}} \frac{\partial }{\partial t}\{\sqrt{\gamma} [ (p + \rho) u_{0}u^{0} - p] \} - \frac{1}{\sqrt{\gamma}} \partial_{i} \{ \sqrt{\gamma} (p + \rho) 
u_{0}u^{i} \} 
\nonumber\\
&&- K_{k}^{\ell} [ (p + \rho) u^{k} u_{\ell} + p\delta_{\ell}^{k}] =0,
\label{con2}
\end{eqnarray}
where $\gamma = {\rm det}(\gamma_{ij})$.

It is useful to recall, from the Bianchi identities, that the intrinsic 
curvature tensor and its trace satisfy the following identity
\begin{equation}
\nabla_{j} r^{j}_{i} = \frac{1}{2} \nabla_{i} r.
\label{fromb}
\end{equation}

Note, finally, that combining  Eq. (\ref{E100}) with the trace 
of Eq. (\ref{E1ij}) the following equation is obtained:
\begin{equation}
{\rm Tr}K^2 + K^2 + r - 2 \dot{K} = 8\pi G( \rho - 3p).
\label{trace}
\end{equation}
Equation (\ref{trace}) allows to re-write Eqs. (\ref{E100}), (\ref{E1ij}) and (\ref{E10i})  as 
\begin{eqnarray}
&& \dot{K} - {\rm Tr} K^2 = 8\pi G\biggl[ (p+ \rho) u_{0}\,u^{0} +
\frac{p -\rho}{2} \biggr], 
\label{E200}\\
&& \frac{1}{\sqrt{\gamma}} \frac{\partial}{\partial t}\biggl(\sqrt{\gamma} \,K_{i}^{j}\biggr) - r_{i}^{j} = 8\pi G \biggl[ - (p+ \rho) u_{i} u^{j} 
+ \frac{p -\rho}{2} \delta_{i}^{j} \biggr],
\label{E2ij}\\
&& \nabla_{i} K - \nabla_{k} K^{k}_{i} = 8\pi G (p + \rho) u_{i}\,u^{0}.
\label{E20i}
\end{eqnarray}

\renewcommand{\theequation}{3.\arabic{equation}}
\section{Anti-newtonian  solutions}
\setcounter{equation}{0}
Consider the situation where the scalar curvarure $r$ and the 
velocities are all much smaller than $K^2$ and $\dot{K}$. 
This situation describes the occurrence of a ``anti-newtonian" 
regime and has been previously investigated by Tomita 
\cite{tom1,tom2} as well as by Deruelle and collaborators \cite{dr1,dr2,dr3} 
interested in various aspects of this approximation. 

If $K^2$ and $\dot{K}$ are both leading in comparison with the 
curvature and velocity contributions, Eqs. (\ref{E200}), (\ref{E2ij}) and  
(\ref{con1}) can  be written, as
\begin{eqnarray}
&& \dot{K} - \frac{K^2}{3} - {\rm Tr} Q^2 = 4\pi G  ( 3 p + \rho),
\label{an1}\\
&& \frac{1}{\sqrt{\gamma}} \frac{\partial}{\partial t} ( \sqrt{\gamma} K) 
 = 12 \pi G  ( p -\rho),
\label{an2}\\
&&  \frac{1}{\sqrt{\gamma}} \frac{\partial}{\partial t}( \sqrt{\gamma} Q_{i}^{j}) =0,
\label{an3}
\end{eqnarray}
where we defined 
\begin{equation}
K_{i}^{j} = \frac{K}{3} \delta_{i}^{j} + Q_{i}^{j},
\label{defQ}
\end{equation}
with $Q = Q_{i}^{i} =0$. 
From Eq. (\ref{con2}), covariant conservation of the energy-momentum 
tensor implies 
\begin{equation}
 \dot{\rho}  + \frac{\dot{\gamma}}{2\gamma}( p + \rho)  =0.
\label{con2a}
\end{equation}
From Eq. (\ref{an3}) and (\ref{con2a}) it follows  
\begin{eqnarray}
&& \rho = \rho_{0}(\vec{x}) \gamma^{- \frac{w+1}{2}},
\label{rhoeq}\\
&& Q_{i}^{j} = \frac{\lambda_{i}^{j}}{\sqrt{\gamma}}
\label{Qeq}
\end{eqnarray}
where $\lambda_{i}^{i} = 0$  and $\lambda_{i}^{j}$ only depends 
on space (and not on time); note that we used the fact that 
$p = w \rho$.

Inserting then Eq. (\ref{an2}) into Eq. (\ref{an1}) to eliminate the 
energy density we obtain:
\begin{equation}
\dot{K} - \frac{w+ 1}{2} K^2 + \frac{3}{4} (w-1) \frac{{\rm Tr} \lambda^2}{\gamma} =0, 
\label{decan}
\end{equation}
where, as usual, ${\rm Tr} \lambda^2 = \lambda_{i}^{j}\lambda_{j}^{i} $.
A new variable can now be defined, namely, 
\begin{equation}
M = \gamma^{\frac{w + 1}{4}};
\label{def}
\end{equation}
recalling that $ \dot{\gamma} = - 2\gamma\,K$, Eq. (\ref{decan}) 
becomes, in terms of $M$:
\begin{equation}
\ddot{M} = \frac{3}{8} (w^2 -1) {\rm Tr}\lambda^2 M^{\frac{w -3}{w + 1}}.
\label{part}
\end{equation}
Integrating once with respect to the cosmic time, the following relation can
be obtained: 
\begin{equation}
\dot{M}^2 = \frac{3}{8} (w + 1)^2 \,{\rm Tr} \lambda^2 M^{\frac{ 2 ( w -1)}{w + 1}}  + N(\vec{x}).
\label{TOM}
\end{equation}
Equation (\ref{TOM}) can be solved (either analytically or numerically) 
for a given value of the barotropic index.
 In particular, in the case $w =0$, recalling Eq. (\ref{def}) we can write:
 \begin{equation}
 \sqrt{\gamma} = N(\vec{x}) [ t - t_{1}(\vec{x})]^2 - \frac{3}{8} \frac{{\rm Tr} \lambda^2}{N(\vec{x})},
 \label{dust1}
 \end{equation}
 where a further integration constant appear. Either $N(\vec{x})$ or $t_{1}(\vec{x})$ can be fixed by exploiting the remaining gauge 
 freedom of the synchronous system. 
 Note that the momentum constraint of Eq. (\ref{E20i}) 
 implies 
 \begin{equation}
 u^{0} u_{i} = \frac{1}{12\pi G \rho (w + 1)} \biggl[ \partial_{i} K - 
 \frac{3}{2\sqrt{\gamma}}  ( \Gamma_{k\ell}^{k} \lambda_{i}^{\ell} - 
 \Gamma_{i m}^{k} \lambda_{k}^{m} )\biggr],
 \end{equation}
 that is of higher order in the gradients, as expected.
 
 The deceleration parameter defined in 
 Eq. (\ref{defq}) can be written, using Eqs. (\ref{defq}) and (\ref{Qeq}) as 
 \begin{equation}
 q(\vec{x}, t) = \frac{1}{4} \biggl(\frac{ 2 \gamma \, K^2 - 3 {\rm Tr} \lambda^2}{
 \gamma\, K^2 + 3 {\rm Tr} \lambda^2}\biggr)
 \end{equation}
i.e., using Eq. (\ref{dust1})
\begin{equation}
q(\vec{x},t) = \frac{1}{4} \biggl(\frac{8 N(\vec{x})^2 [t  -t_1(\vec{x})]^2  - 3 {\rm Tr} \lambda^2}{4 N(\vec{x})^2 [t - t_1(\vec{x})]^2 + 3 {\rm Tr}\lambda^2}\biggr),
\label{qdustcalc}
\end{equation}
whose large time limit is $1/2$. Note that by looking superficially 
at Eq. (\ref{qdustcalc}) one would be tempted to conclude that  
$q$ might be negative. This is not correct since the numerator in
of the expression in Eq. (\ref{qdustcalc}) is exactly proportional to $\sqrt{\gamma}\geq 0$.

Now one could choose, as seed metric, the solution found with this method.
Then compute the spatial curvature and obtain the following order in the gradient expansion and  so on \cite{dr1,dr3}. However, the 
solution of Eq. (\ref{TOM}) cannot be inverted analytically for a generic 
barotropic index $w$.  

\renewcommand{\theequation}{4.\arabic{equation}}
\section{Quasi-isotropic solution and gradient expansion}
\setcounter{equation}{0}

The solutions of Eqs. (\ref{E100})--(\ref{E10i}) can be classified, according 
to their degree of isotropy, in quasi-isotropic solutions 
and fully anisotropic solutions. For instance, close to the initial 
(big-bang) singularity the solution of Eqs. (\ref{E100})--(\ref{E10i}) 
is, in general, fully anisotropic. In the opposite limit, i.e. 
far from the initial singularity, the possibility of quasi-isotropic 
solutions becomes more relevant. Quasi-isotropic solutions 
exist, indeed, only in the presence of matter \cite{lif1,lif2} (see also \cite{BK1,KL,BK2}). 

Let us now look for solutions of the previous system of equations in the form 
of a gradient expansion. 
In other words we shall be considering $\gamma_{ij}$ written in the form 
\begin{equation}
\gamma_{ik} = a^2(t) [ \alpha_{ik}(\vec{x}) + \beta_{ik}(t,\vec{x})],\,\,\,\,\,\,\,\,\,\,
\gamma^{kj} = \frac{1}{a^2(t)} [ \alpha^{kj} - \beta^{kj} (t, \vec{x})],
\label{exp}
\end{equation}
where $\beta(\vec{x},t)$ is considered to be the first-order correction 
in the spatial gradient expansion. Note that from Eq. (\ref{exp}) 
$\gamma_{ik}\gamma^{kj} =  \delta_{i}^{j} + {\cal O}(\beta^2)$.
The logic is now very simple since Einstein equations will determine 
the specific form of $\beta_{ij}$ once the specific form of $\alpha_{ij}$ 
is known. 

Using Eq. (\ref{exp}) into Eqs. (\ref{extr}) we obtain
\begin{equation}
K_{i}^{j} = - \biggl( H \delta_{i}^{j} + \frac{\dot{\beta}_{i}^{j}}{2} \biggr),
\qquad K = - \biggl( 3 H + \frac{1}{2}\dot{\beta}\biggr),
\qquad {\rm Tr} K^2 = 3 H^2 + H \dot{\beta},
\end{equation}
where, with obvious notations, $ H = \dot{a}/a$.

From Eq. (\ref{E10i}) it also follows  
\begin{equation}
\nabla_{k} \dot{\beta}_{i}^{k} - \nabla_{i} \dot{\beta} = 16\pi G u_{i} \,\,u^{0} (p+ \rho).
\label{momex}
\end{equation}
The explicit form of the momentum constraint suggests to 
look for the solution in a separable form, namely, 
 $\beta_{i}^{j}(t,\vec{x}) = g(t) {\cal B}_{i}^{j}(\vec{x})$. Thus  
Eq. (\ref{momex}) becomes 
\begin{equation}
\dot{g} (\nabla_{k} {\cal B}^{k}_{i} - \nabla_{i} {\cal B}) = 16 \pi G u_{i} u^{0} (p + \rho).
\label{momex1}
\end{equation}
Using this parametrization and solving the constraint for $u_{i}$, Einstein
equations to second order in the gradient expansion reduce then 
to the following equation:
\begin{equation}
( \ddot{g} + 3 H \dot{g}) {\cal B}_{i}^{j} + H \dot{g} {\cal B} \delta_{i}^{j} + 
\frac{2 {\cal P}_{i}^{j}}{a^2} = \frac{w -1}{3 w + 1} (\ddot{g} + 2 H \dot{g}) B 
\delta_{i}^{j}.
\label{dec}
\end{equation}
In Eq. (\ref{dec}) the spatial curvature tensor has been parametrized as 
\begin{equation}
r_{i}^{j} = \frac{{\cal P}_{i}^{j}}{a^2}.
\label{defr}
\end{equation}
Recalling that 
\begin{equation}
H= H_{0} a^{- \frac{3( w + 1)}{2}}, \qquad \dot{H} = - \frac{3 ( w + 1)}{2} 
H^2,
\end{equation}
the solution for Eq. (\ref{dec}) can be written as
\begin{eqnarray}
&& {\cal B}_{i}^{j} = - \frac{4}{H_{0}^2 ( 3 w + 1) ( 3 w + 5)} \biggl( {\cal P}_{i}^{j} - \frac{ 5 + 6 w - 3 w^2}{4 ( 9 w + 5)}{\cal P} \delta_{i}^{j} \biggr),
\nonumber\\
&& {\cal B} = - \frac{\cal P}{H_{0}^2 ( 9 w + 5)},
\label{calB}
\end{eqnarray}
with  $g(t)$ simply given by 
\begin{equation}
g(t) = a^{3 w +1}.
\end{equation}
Note that, in Eq. (\ref{calB}), $H_{0} = 2/[ 3 (w + 1)\, t_0]$. Eqs. (\ref{calB}) 
agree with the expression obtained in the case $w =1/3$ in Ref. \cite{LK1} (where the notations are such that $t_0 = 1$).
Equation (\ref{calB}) can be also inverted, i.e. ${\cal P}_{i}^{j}$ can
be easily expressed in terms of ${\cal B}_{i}^{j}$ and ${\cal B}$:
\begin{equation}
{\cal P}_{i}^{j} = - \frac{H_{0}^2}{4} [ {\cal B} \delta_{i}^{j} ( 6w + 5 - 3 w^2) 
+ {\cal B}_{i}^{j} ( 3 w + 5) ( 3 w + 1) ]
\end{equation}
 
Using Eq. (\ref{fromb}) 
the peculiar velocity field and the energy density can also be written as 
\begin{eqnarray}
&& u^{0} u_{i} = - \frac{3}{8\pi G \rho} \biggl(\frac{w}{3 w+ 5}\biggr)a^{3 w + 1} H \partial_{i} {\cal B}(\vec{x}),
\nonumber\\
&& \rho = \frac{3 H_{0}^2}{8\pi G}\biggl[ a^{-3( w + 1)} 
- \frac{w +1}{2} {\cal B}(\vec{x}) a^{-2} \biggr].
\end{eqnarray}
Let us therefore rewrite the solution in terms of $\gamma_{ij}$, i.e. 
\begin{equation}
\gamma_{i j} = a^2(t)[ \alpha_{i j} (\vec{x}) + \beta_{ij} (\vec{x}, t) ] = 
a^{2}(t)\biggl[\alpha_{ij}(\vec{x}) +  a^{ 3 w + 1} {\cal 
B}_{i j} (\vec{x}) \biggr].
\label{qisol}
\end{equation}
Concerning this solution a few comments are in order:
\begin{itemize}
\item{} if $w > -1/3$, $\beta_{ij}$ becomes large as 
$a\to \infty$ (note that if $w= - 1/3$, $a^{3 w + 1}$ is constant);
\item{} if $w < - 1/3$, $\beta_{ij}$  vanishes as  $a\to \infty$;
\item{} if $w < -1$, $\beta_{ij}$ not only the gradients become sub-leading 
but the energy density also increases as $a\to \infty$.
\item{} to the following order in the perturbative expansion the time-dependence is easy to show: $\gamma_{ij} \simeq  a^2(t)[\alpha_{ij} + a^{3w + 1} {\cal B}_{ij} +a^{2(3w + 1)}  {\cal E}_{ij}]$ and so 
on for even higher order terms;
clearly the calculation of the curvature tensors will now be just a bit 
more cumbersome.
\end{itemize}
The first two comments are rather elementary. The first comment 
simply expresses the fact that if $w > -1/3$ the gradients decay 
close to the initial big-bang singularity (but not far from it!).  The second 
comment simply justifies why during a phase dominated by 
an effective cosmological constant (or by a fluid violating 
the strong energy condition) the gradients are washed out.
The third comment may have some interesting implications 
for the study of big rip singularities or for the more general case of sudden \cite{barrow1,barrow2,barrow3}  singularities.  It would be for instance 
interesting to investigate if, in general, sudden singularities 
will exhibit or not some type of BKL oscillations 
that are known to be present in the case of the initial big-bang singularity.

\renewcommand{\theequation}{5.\arabic{equation}}
\section{Equivalent forms of gradient expansion}
\setcounter{equation}{0}
Consider now the following parametrization of the perturbed 
metric:
\begin{equation}
\gamma_{i j} = e^{-2\Psi(t,\vec{x})} a^2(t) [\delta_{i j} +\mu_{ij}(\vec{x}, t)],
\label{sup}
\end{equation}
where $\mu_{i}^{i}=0$. This parametrization has been used 
in \cite{K} and in \cite{mgio3}. Equation (\ref{sup}) may seem, superficially,
different from the one of Eq. (\ref{exp}) since, on one hand, the analog 
of $\alpha_{ij}$ has a very specific tensor structure
coinciding with the three-dimensional 
Kroeneker symbol; on the other hand $\Psi$ is allowed to depend both on space and time. 

Equation (\ref{sup})
\begin{equation}
K_{i}^{j} = (\dot{\Psi} - H) \delta_{i}^{j} - \frac{\dot{\mu}_{i}^{j}}{2}.
\label{Kij}
\end{equation}
In this case, the Hamiltonian constraint of Eq. (\ref{E100}) leads to 
\begin{equation}
6(\dot{\Psi} - H)^2 + \frac{e^{2 \Psi}}{a^2}\biggl[ 4 \nabla^2 \Psi - 2 (\nabla\Psi)^2] = 16\pi G 
\biggl[ \rho + (p + \rho) \frac{e^{2\Psi}}{a^2} (u^2 - 
\tilde{u}^2)\biggr],
\end{equation}
where $u^2 = u_{i} u_{j} \delta^{ij}$ and $\tilde{u}^2 = u_{i} u_{j} \mu^{i j}$.
From the $(ij)$ components of Einstein equations we get, after linear combination, 
\begin{eqnarray}
&& 3 (\ddot{\Psi} - \dot{H}) - 3 (\dot{\Psi} - H)^2 = 4\pi G ( \rho + 3 p) +
8\pi G (p + \rho) \frac{e^{2\Psi}}{a^2} ( u^2 - \tilde{u}^2),
\nonumber\\
&& \ddot{\mu}_{i}^{j} + 3 H \dot{\mu}_{i}^{j} = 
-2 \frac{ e^{2 \Psi}}{a^2 } \biggl[\partial_{i} \partial^{j} \Psi - \frac{1}{3} 
\nabla^2 \Psi \delta_{i}^{j}+ \partial_{i}\Psi \partial^{j}\Psi - 
\frac{1}{3} (\nabla\Psi)^2 \delta_{i}^{j} \biggr]
\nonumber\\
&& + 16\pi G \,( p +\rho) \frac{e^{2\Psi}}{a^2} \biggl[ ( u_{i} u^{j} 
- \tilde{u}_{i} \tilde{u}^{j} ) - \frac{u^2 - \tilde{u}^2}{3} \delta_{i}^{j}\biggr],
\end{eqnarray} 
where $\nabla^2 \Psi = \delta^{ij} \partial_{i} \partial_{j} \Psi$ and where 
$(\nabla\Psi)^2 = \delta^{ij} \partial_{i} \Psi\partial_{j} \Psi$.
Finally, the momentum constraint, to first-order in the gradient expansion, is 
\begin{equation}
\partial_{i} \dot{\Psi}  + \frac{1}{4} \partial_{k} \dot{\mu}_{i}^{k}= 4\pi G (p + \rho) u_{i} u^{0}.
\label{mompsi}
\end{equation}
A more tractable form of the system can be obtained by eliminating 
the energy density; hence the following pair of equations 
emerges:
\begin{eqnarray}
&& - \ddot{\Psi} + \frac{ 3( w + 1)}{2} \dot{\Psi}^2 - 3 H ( w+ 1) \dot{\Psi}
+ \frac{ 3 w + 1}{3 a^2} e^{ 2 \Psi} \biggl[ \nabla^2 \Psi - \frac{1}{2} (\nabla\Psi)^2 \biggr] =0,
\label{pseq}\\
&& \ddot{\mu}_{i}^{j} + 3 H \dot{\mu}_{i}^{j} = -\frac{2}{a^2} e^{2\Psi}
\biggl[ \partial_{i}\partial^{j}\Psi - \frac{1}{3} (\nabla^2\Psi)\delta_{i}^{j} + 
\partial_{i}\Psi \partial^{j} \Psi - \frac{1}{3}(\nabla\Psi)^2 \delta_{i}^{j} \biggr].
\label{mueq}
\end{eqnarray}
The solution of Eqs. (\ref{pseq}) and (\ref{mueq}) can be written as:
\begin{eqnarray}
&& \Psi = f + \frac{j_1(w)}{H_{0}^2} a^{ 3 w + 1} e^{ 2 f} \biggl[ \nabla^2 f - \frac{1}{2} (\nabla f)^2\biggr],
\label{solpsi}\\
&& \mu_{i}^{j} = -\frac{j_2(w)}{H_{0}^2} 
 a^{ 3 w + 1} e^{2 f} \biggl[ \partial_{i} \partial^{j} f - \frac{1}{3} (\nabla^2 f) \delta_{i}^{j} 
+ \partial_{i} f \partial^{j} f - \frac{1}{3} (\nabla f)^2 \delta_{i}^{j} \biggr].
\label{solmu}
\end{eqnarray}
where 
\begin{equation}
j_{1}(w) = \frac{2}{3 ( 9 w + 5)}, \qquad j_{2}(w) = \frac{4}{( 3 w + 1) ( 3 w+ 5)}
\end{equation}
and where $f(\vec{x})$ is an arbitrary function depending upon space but 
not upon time. In this function is encoded the information on the specific 
kind of super-horizon fluctuations. The velocity fields have been consistently 
neglected since they are of higher order in the gradient expansion. 
This aspect can be appreciated by inserting the solution (\ref{solpsi}) 
into Eq. (\ref{mompsi}): $\partial_{i} \dot{\Psi}$  only receives 
contribution from the second term in Eq. (\ref{solpsi}) (i.e. the one 
proportional to $j_1(w)$). Thus $u_{i}$ will contain three gradients 
and will be negligible at this order. Notice, however, that the contribution 
of the velocity field must be necessarily taken into account when going 
to orders higher than the second in the gradient expansion.

Since Eq. (\ref{solpsi}) and (\ref{solmu}) 
are derived under the approximation that terms with more than two gradients 
are neglected,  the terms $\dot{\Psi}^2$ has to be negligible 
with respect to $\ddot{\Psi}$, i.e. 
\begin{equation}
2|\ddot{\Psi}| \gg 3(w + 1)|\dot{\Psi}|^2.
\label{condap}
\end{equation}
This aspect can be appreciated by computing $\dot{\Psi}^2$ from Eq. 
(\ref{solpsi}): the only contribution to $\dot{\Psi}^2$ comes
from the term proportional to $j_1(w)$ in Eq. (\ref{solpsi}) and this 
has four spatial gradients.
A relevant consequence of this basic observation 
is that when computing ${\rm Tr} K^2$ from Eq. (\ref{Kij})
the correct result, within the approximations made so far, is 
\begin{equation}
{\rm TrK^2} = H^2 - 2 H \dot{\Psi} \geq 0
\label{inequality}
\end{equation}
and not ${\rm Tr}K^2 = H^2 - 2H \dot{\Psi} + \dot{\Psi}^2$. Notice that the 
inequality appearing in Eq. (\ref{inequality}) is a direct 
consequence of the inequalities reported in Eq. (\ref{trin}): 
${\rm Tr}K^2 $ is {\em always} (i.e. at any order) positive semi-definite. 
This simple observation has a simple consequence, i.e. that 
\begin{equation}
\dot{\Psi} \leq \frac{H}{2}.
\label{conda}
\end{equation}
For future convenience, let us rewrite the solution of Eq. (\ref{solpsi})
as 
\begin{equation}
\Psi(\vec{x},t) = f(\vec{x}) + a^{3w + 1} \overline{\Psi}_{0}(\vec{x})
\label{Psi0}
\end{equation}
where $\overline{\Psi}_{0}$ can be read-off from Eq. (\ref{solpsi})
\begin{equation}
\overline{\Psi}_{0}(\vec{x}) = \frac{j_1(w)}{H_{0}^2} e^{ 2 f} \biggl[ \nabla^2 f - \frac{1}{2} (\nabla f)^2\biggr].
\label{defpsi0}
\end{equation}
From Eq. (\ref{Psi0}), by taking the first time derivative, 
 it follows easily that $\dot{\Psi} = ( 3 w + 1)a^{3 w+ 1}H \overline{\Psi}_{0}$, and this implies: inserting the obtained expression for 
 $\dot{\Psi}$ into Eq. (\ref{conda}), the inequality 
 puts a condition on $\overline{\Psi}_{0}$, i.e. 
 \begin{equation}
 a^{3 w+1} |\overline{\Psi}_{0}| \leq \frac{1}{2(3w + 1)},
 \label{condb}
 \end{equation}
  that translates, in the case $w = 0$ into 
\begin{equation}
  a \overline{\Psi}_{0} \leq \frac{1}{2}.
\label{condb0}
\end{equation}
 
We are now in condition to clarify the relation 
between the perturbative expansion discussed in the present 
Section and the general quasi-isotropic gradient expansion 
derived in Eqs. (\ref{calB}) and (\ref{qisol}). The conclusion
will be that, if the positivity of ${\rm Tr}K^2$ is enforced, the 
two expansions are exactly equivalent, in the sense 
that the expansion discussed in the present Section is just 
as sub-case of the general quasi-isotropic expansion.

For this purpose let us then consider Eq. (\ref{qisol}) and let use the freedom 
to specify $\alpha_{ij}$ in Eq. (\ref{qisol}) and let us choose
\begin{equation}
\alpha_{ij } (\vec{x}) = e^{ - 2 f (\vec{x})} \delta_{ij}.
\label{specific}
\end{equation}
Then, the spatial curvature tensor can be immediately 
computed using Eq. (\ref{specific}) and  Eq. (\ref{intr}). Thus, recalling 
Eq. (\ref{defr}), ${\cal P}_{i}^{j}$ can be easily computed: 
\begin{equation}
{\cal P}_{i}^{j} = e^{2 f} \biggl[ \nabla^2 f \delta_{i}^{j} + \partial_{i} \partial^{j} f
- (\nabla f)^2 \delta_{i}^{j} + \partial_{i}f \partial^{j} f\biggr].
\end{equation}
Consequently, from Eq. (\ref{calB}), the tensor ${\cal B}_{ij}$ will be
\begin{equation}
{\cal B}_{i}^{j} = - \frac{4 e^{2 f}}{H_{0}^2 ( 3 w +1 ) ( 3 w + 5)} \biggl\{ 
\biggl[ \frac{ 3 w ( w + 1) }{ ( 9 w + 5)} \nabla^2 f - 
\frac{ 3 w^2 + 12 w + 5}{ 2( 9 w + 5)} ( \nabla f)^2 \biggr] \delta_{i}^{j} 
+ \partial_{i} \partial^{j} f + \partial_{i} f \partial^{j} f \biggr\}.
\label{expl}
\end{equation}
This expression is still difficult to compare with the parametrization employed 
in the present section since ${\cal B}_{i}^{j}$ is not traceless. Let us then 
separate the traceless contribution by writing  $\gamma_{ij}(\vec{x}, t)$ 
as 
\begin{equation}
\gamma_{i j}(\vec{x}, t) = a^{2}(t) e^{- 2 \tilde{\Psi}} [ \delta_{i j} + \tilde{\mu}_{i j} (\vec{x}, t) ]
\end{equation}
where  $\tilde{\mu}_{i}^{i} =0$. The traceless tensor $\tilde{\mu}_{i j}$ is essentially 
given by the traceless part of $\beta_{ij} = a^{(3w + 1)} {\cal B}_{ij}$ where 
${\cal B}_{ij}$ is given by Eq. (\ref{expl}) . Therefore, 
 we can also write the solution in terms of $\tilde{\Psi}$:
\begin{equation}
 \tilde{\Psi} = f(\vec{x}) - \frac{1}{2} \ln{\biggl\{ 1 - 
\frac{4}{3} \frac{e^{ 2 f}}{( 9 w + 5) H_{0}^2} a^{ 3 w + 1} \biggl[  \nabla^2 f -
\frac{1}{ 2} (\nabla f)^2 \biggr]\biggr\}}.
\label{genres}
\end{equation}
In the specific case $ w = 0$ the previous expression becomes 
\begin{equation}
\tilde{\Psi} = f(\vec{x}) - \frac{1}{2} \ln{\biggl\{\biggl[ 1 - 
\frac{4}{15 H_{0}^2} e^{ 2 f} a \biggl[  \nabla^2 f - 
\frac{ 1 }{ 2 } (\nabla f)^2 \biggr]\biggr\} }.
\label{genres0}
\end{equation}
Let us now compare Eqs. (\ref{genres}) and (\ref{genres0}) with 
Eq. (\ref{solpsi}). For for sake of simplicity let us consider the case 
$w=0$. Therefore, in this case, $\tilde{\Psi}$ of Eq. (\ref{genres0}) 
can be written as 
\begin{equation}
\tilde{\Psi} = f(\vec{x}) - \frac{1}{2}\ln{[1 - 2 a \overline{\Psi}_{0}]},
\label{log1}
\end{equation}
where $\Psi_{0}(\vec{x})$ is the same one defined in Eq. (\ref{defpsi0}) and it includes the contribution of the spatial gradients.
Recalling now Eq. (\ref{condb0}) we clearly see that the 
argument of the logarithm appearing in Eq. (\ref{log1}) is {\em only}
compatible with ${\rm Tr}K^2 $ being positive semidefinite, if
$ 2 a \overline{\Psi}_{0} \leq 1$. This implies 
that, as anticipated, $\tilde{\Psi} = \Psi$ within the accuracy 
of the approximation.  

The condition coming from ${\rm Tr} K^2 \geq 0$ is necessary 
for the consistency of the approximation of the present section. It 
is however not sufficient. This means that an even more restrictive 
condition on $\overline{\Psi}_{0}$ stems from Eq. (\ref{condap}). 
Indeed, using the parametrization (\ref{Psi0}) it follows, in the case 
$w=0$, that 
\begin{equation}
a |\overline{\Psi}_{0}| \leq \frac{1}{3}.
\label{condc}
\end{equation}
This inequality follows directly from Eq. (\ref{condap}) by 
recalling, from Eq. (\ref{Psi0}), that,  for $w=0$, 
$\Psi = f + a \overline{\Psi}_{0}$. Thus $\ddot{\Psi}= (\dot{H} + H^2) 
|\overline{\Psi}_{0}|a$; using the last expression in Eq. (\ref{condap})
and recalling that $2\dot{H} = - 3 H^2$ (in the case of dusty matter) we 
Eq. (\ref{condc}) is readily obtained. As anticipated the condition 
(\ref{condc}) is more restrictive than the one derived in (\ref{condb0})
always in the $w=0$ case. In fact, Eq. (\ref{condc}) implies 
$-1/3 \leq a\overline{\Psi}_{0} \leq 1/3$ while Eq. (\ref{condb0}) 
implies only that $ a \overline{\Psi}_{0} < 1/2$.  

The last step of our discussion is to show that not only $\tilde{\Psi} =\Psi$
but also that $\mu_{ij} = \tilde{\mu}_{ij}$. This is immediate recalling 
that $\tilde{\mu}_{ij}$ is constructed from the traceless part of $\beta_{ij}$ 
whose tensor structure in given, in terms of the intrinsic curvature tensor, 
by Eq. (\ref{calB}). Bearing  now in mind Eq. (\ref{specific}), it is immediate 
to show that $\tilde{\mu}_{ij}$ equals $\mu_{ij}$ given in Eq. (\ref{solmu}).

\renewcommand{\theequation}{6.\arabic{equation}}
\section{Deceleration parameter(s)}
\setcounter{equation}{0}
In this section we will be concerned mainly with 
the calculation of the deceleration parameter.
In particular we want to show that the deceleration 
parameter for a matter-dominated Universe can never become 
negative. In a complementary perspective 
one could say that if the deceleration parameter 
becomes negative, within the gradient expansion
discussed in the present paper, the 
conditions of validity of the perturbative expansion 
must be violated.  

Let us therefore start from the general definition of the 
deceleration parameter valid in the inhomogeneous case 
and reported in Eq. (\ref{defq}). Then let us consider the 
parametrization of Eq. (\ref{Psi0}) and in the case $w=0$.
Inserting Eq. (\ref{Psi0}) into Eq. (\ref{defq}) we obtain 
\begin{equation}
q(\vec{x},t) = -1 + \frac{3/2 - a \overline{\Psi}_{0}/2}{ ( 1 - 2 a \overline{\Psi}_{0})} = \frac{3 a \overline{\Psi}_{0}  + 1}{2 ( 1 - 2 a \overline{\Psi}_{0} )}.
\label{correctq}
\end{equation}
In order to derive this result, we inserted Eq. (\ref{Psi0}) 
into $K$, $\dot{K}$ and ${\rm Tr}K^2$ appearing in the 
general definition (\ref{defq}). Then we used the relation 
$2\dot{H} = - 3H^2$ (valid in the dust-dominated case). 
Finally, we consistently neglected the terms $\dot{\Psi}^2$ 
that, as explained in the previous Section, are of higher order 
since they contain $4$ gradients. This means, in practice, that 
${\rm Tr}K^2 = H^2 ( 1 - 2 a\overline{\Psi}_{0})$. 
Concerning Eq. (\ref{correctq}) the following comments are in order:
\begin{itemize}
\item{}  as derived in the previous Section the validity of the perturbative expansion implies in the case $w=0$ that $a|\overline{\Psi}_{0}| \leq 1/3$
(see Eq. (\ref{condc}) and comments therein);
\item{} therefore, it is clear from the expression of reported in the 
second equality of Eq. (\ref{defq}) that $q$ is {\em always} positive 
semi-definite  for $-1/3 \leq a\overline{\Psi}_{0} \leq 1/3$;
\item{} furthermore, when $a \overline{\Psi}_{0} \to -1/3$ 
(corresponding to the maximally underdense Universe allowed 
by the validity of the perturbative expansion), then $q\to 0$.
\end{itemize}
The last occurrence corresponds to a gradient-dominated Universe. This 
result was already derived in \cite{mgio3}. 
Therefore, the validity of the perturbative expansion forbids a negative 
deceleration parameter and, therefore, the acceleration of an Universe 
dominated by dusty matter (i.e. $w =0$).

Notice also that
Eq. (\ref{condc}) implies an upper limit on $a(t)$, i.e. 
\begin{equation}
a \leq \frac{1}{3 |\overline{\Psi}_{0}|}.
\label{bound}
\end{equation}
If sufficiently small values of  $\overline{\Psi}_{0}$ are allowed, $a$ can be 
rather large. To support their argument, the authors of Ref. \cite{K} took 
(in their figure) the smallest value of $\Psi_{0}$ to be 
 $\overline{\Psi}_{0} \sim -1/4$. This simply means, according 
 to Eq. (\ref{bound}) that   $a \leq \frac{4}{3}$.  
 In light of the discussion of the present paper the soundness of this 
 results stems directly from the equivalence 
 of the gradient expansions discussed in Sections 4 and 5. 
 In fact, it is clear that the quasi-isotropic gradient expansion 
 of Section 4 is suitable in the limit $a\to 0$, i.e. in the limit 
 where it was correctly applied in the early sixties.
 
 From the results of Section 4 and 5, moreover, it is 
 clear that there are cases when the gradient expansion 
 may be suitable to describe the large times behaviour. But 
 these cases are precisely the ones for which $w < -1/3$. In this 
 class of fluid models, as remarked in Section 4, the contribution 
 of the gradients becomes progressively less important as the 
 Universe expands. Unfortunately, the cases $w <-1/3$  correspond 
 to an Universe that is not dust-dominated.
 
 Finally, in order to complete our discussion, let us see what happens
 if we include terms of higher order in the gradient expansion and take the
 limit for $a \to \infty$. In this case we have that ${\rm Tr}K^2 = 
 (H- \dot{\Psi})^2$. Then using the parametrization of Eq. (\ref{Psi0}) 
 into Eq. (\ref{defq}) we do find that the deceleration parameter 
 is given by
 \begin{equation}
 q(\vec{x},t) = -1 + \frac{3/2 - a \overline{\Psi}_{0}/2}{ ( 1 -  a \overline{\Psi}_{0})^2}.
 \label{funny}
 \end{equation}
 The conclusion of Ref. \cite{K} is that if we extend the validity of Eq. (\ref{funny}) into the far future, the 
 deviation of the deceleration parameter from $q=1/2$ becomes larger 
 and larger approaching the asymptotic value $q=-1$.
 On the basis of the gradient expansion discussed in the present paper
 such a chain of arguments is arbitrary. 
 
\renewcommand{\theequation}{7.\arabic{equation}}
\section{Concluding remarks}
\setcounter{equation}{0}
In the present paper the gradient expansion has been discussed 
for a generic perfect fluid with barotropic index $w$. It has been 
shown that formally different realizations of the gradient expansion
are in fact equivalent if the perturbative expansion is correctly 
handled. The aim of the present analysis was to understand recent claims suggesting that perfect fluids 
with $w\geq 0$ can act as effective dark-energy candidates. The results 
of the present study imply that if $0\leq w \leq 1 $ the deceleration parameter is always positive semi-definite. In the present analysis we kept terms 
up to two gradients in the inhomogeneous expansion. At this 
order, the negativity of the deceleration parameter for matter sources with 
$0\leq w \leq 1 $ is just a signal of the breakdown of the perturbative expansion.  The techniques discussed in the present paper may also 
be relevant in order to study a possible inhomogeneous (and anisotropic )
approach of big rips and sudden singularities. We leave 
this problem for future investigations.

\vskip 2cm
{\em Note added}.  After completing the present paper we became aware of 
astro-ph/0506534 by E. Kolb, S. Matarrese and A. Riotto. These 
authors study solutions of the gradient expansion discussed in Section 
5 of the present paper.  Their results are less general since 
they only consider a dust-dominated Universe. We checked that our solutions 
coincide with their solutions in the limit $w=0$ (our $ f(\vec{x})$ becomes, in their notations, $5/3 \varphi(\vec{x})$). They also claim that the perturbative expansion they use (i.e. the one we discussed in Section 5) sensibly differs from the gradient expansion we discussed in Section 4 of the present paper.
Our results show the opposite.


\newpage

\begin{thebibliography}{99}

\bibitem{riess1} A.~G.~Riess {\it et al.}  [Supernova Search Team Collaboration],  Astrophys.\ J.\  {\bf 607}, 665 (2004).

\bibitem{tonry1} J.~L.~Tonry {\it et al.}  [Supernova Search Team Collaboration],  Astrophys.\ J.\  {\bf 594}, 1 (2003).

\bibitem{riess2} A.~G.~Riess {\it et al.}  [Supernova Search Team Collaboration],  Astron.\ J.\  {\bf 116}, 1009 (1998).

\bibitem{perlmutter}  S.~Perlmutter {\it et al.}  [Supernova Cosmology Project Collaboration],
  Astrophys.\ J.\  {\bf 517}, 565 (1999).
  
 \bibitem{wmap}  D. N. Spergel {\it et al}., Astrophys. \ J.\ {\bf 148}, 175 (2003).
 
 \bibitem{q1} R. R. Caldwell, R. Dave, and P. J. Steinhardt, Phys. Rev. Lett.  {\bf 80}, 1582 (1998).
 
 \bibitem{q1a} P.~J.~E.~Peebles and B.~Ratra, Astrophys.\ J.\  {\bf 325}, L17 (1988).
 
 \bibitem{q2} S.~M.~Carroll,  Phys.\ Rev.\ Lett.\  {\bf 81}, 3067 (1998).
 
 \bibitem{q3}  P.~J.~E.~Peebles and A.~Vilenkin,  Phys.\ Rev.\ D {\bf 59}, 063505 (1999).
 
 \bibitem{mg1}  M.~Giovannini, Class.\ Quant.\ Grav.\  {\bf 16}, 2905 (1999).
 
 \bibitem{mg2} M.~Giovannini, Phys.\ Rev.\ D {\bf 60}, 123511 (1999).
 
 \bibitem{mg3} D.~Babusci and M.~Giovannini,  Phys.\ Rev.\ D {\bf 60}, 083511 (1999).

\bibitem{padma} T.~Padmanabhan,  Phys.\ Rept.\  {\bf 380}, 235 (2003)

\bibitem{ratra}  P.~J.~E.~Peebles and B.~Ratra,  Rev.\ Mod.\ Phys.\  {\bf 75}, 559 (2003).

\bibitem{rip}   R.~R.~Caldwell, M.~Kamionkowski and N.~N.~Weinberg,   Phys.\ Rev.\ Lett.\  {\bf 91}, 071301 (2003).
  
 \bibitem{rip2} S.~Nojiri, S.~D.~Odintsov and S.~Tsujikawa, Phys.\ Rev.\ D {\bf 71}, 063004 (2005). 
  
\bibitem{visc1}S.~Nojiri and S.~D.~Odintsov, arXiv:hep-th/0505215.

\bibitem{ras}  S.~Rasanen, JCAP {\bf 0411}, 010 (2004).

\bibitem{barrow1} J.~D.~Barrow, Class.\ Quant.\ Grav.\  {\bf 21}, L79 (2004).

\bibitem{barrow2} J.~D.~Barrow, Class.\ Quant.\ Grav.\  {\bf 21}, 5619 (2004).

\bibitem{barrow3}  J.~D.~Barrow and C.~G.~Tsagas, Class.\ Quant.\ Grav.\  {\bf 22}, 1563 (2005).

\bibitem{LK1} E. M. Lifshitz and I. M. Khalatnikov, Sov. Phys. Usp. {\bf 6},  495 (1964).

\bibitem{BLK} V. A. Belinskii, E. M. Lifshitz and I. M. Khalatnikov, Sov. Phys. Usp. {\bf 13}, 745 (1971).

\bibitem{K} E.~Kolb, S.~Matarrese, A.~Notari and A.~Riotto,
  arXiv:hep-th/0503117.

\bibitem{GCA} G.~Geshnizjani, D.~J.~H.~Chung and N.~Afshordi,  arXiv:astro-ph/0503553.

\bibitem{Fla}E.~Flanagan, arXiv:hep-th/0503202.

\bibitem{selhir} C.~M.~Hirata and U.~Seljak, arXiv:astro-ph/0503582.

\bibitem{alnes}  H.~Alnes, M.~Amarzguioui and O.~Gron, arXiv:astro-ph/0506449.

\bibitem{RAS}  S.~Rasanen, arXiv:astro-ph/0504005.

\bibitem{siegel} E.~Siegel and J.~Fry, arXiv:astro-ph/0504421.

\bibitem{mgio3} M. Giovannini,  hep-th/0505222 (CERN-PH-TH/2005-085).

\bibitem{v1} J. D. Barrow, Mon. Not. R. Astr. Soc. {\bf 178}, 625 (1977); {\bf 179}, 47 (1977).

\bibitem{v2} M. Giovannini,
  Phys.\ Rev.\ D {\bf 70}, 103509 (2004);  Class.\ Quant.\ Grav.\  {\bf 22}, 363 (2005).

\bibitem{v3} T.~J.~Battefeld and R.~Brandenberger,
  Phys.\ Rev.\ D {\bf 70}, 121302 (2004).

\bibitem{v4} T.~J.~Battefeld and D.~A.~Easson, Phys.\ Rev.\ D {\bf 70}, 103516 (2004).

\bibitem{tom1} K. Tomita, Prog. Theor. Phys.  {\bf 67}, 1076 (1982).

\bibitem{tom2} K. Tomita, Phys. Rev. D {\bf 48}, 5634 (1993).

\bibitem{dr1}  N. Deruelle and D. Goldwirth, Phys. Rev. D {\bf 51}, 1563 (1995).

\bibitem{dr2} N. Deruelle and K. Tomita, Phys. Rev. D {\bf 50}, 7216 (1994).

\bibitem{dr3} G. Comer, N. Deruelle, D. Langlois, and J. Parry, Phys. Rev. D 
{\bf 49}, 2759 (1994).

\bibitem{lif1} E. M. Lifshitz and I. M. Khalatnikov, Sov. Phys. JETP {\bf 12}, 108 (1960).

\bibitem{lif2} E. M. Lifshitz and I. M. Khalatnikov, Sov. Phys. JETP {\bf 12}, 558 (1961).

\bibitem{BK1} V. A. Belinskii and I. M. Khalatnikov, Sov. Phys. JETP {\bf 30}, 1174 (1970).

\bibitem{KL} I. M. Khalatnikov and E. M. Lifshitz, Phys. Rev. Lett. {\bf 24}, 
76 (1970).

\bibitem{BK2} V. A. Belinskii and I. M. Khalatnikov, Sov. Phys. JETP {\bf 36}, 591 (1973).

\end{thebibliography}
\end{document}